\documentclass{aastex61}

\newcommand{\Msun}{M$_{\odot}$}
\newcommand{\G}{{\it Gaia}}
\newcommand{\GD}{{\it Gaia}-DR2}
\newcommand{\kms}{km\,s$^{-1}$}
\newcommand{\GDT}{{\it Gaia}-DR2-Tycho-2}
\newcommand{\masyr}{mas\,yr$^{-1}$}
\newcommand{\Teff}{$T_{eff}$}

\received{\today}
\revised{\today}
\accepted{\today}
\submitjournal{AJ}

\shorttitle{Bright binary and multiple stars from {\it Gaia}-DR2 and VO}
\shortauthors{Jim\'enez-Esteban et al.}

\begin{document}

\title{A catalog of wide binary and multiple systems of bright stars from {\it Gaia}-DR2 and the Virtual Observatory}


\correspondingauthor{F. M. Jim\'enez-Esteban}
\email{fran.jimenez-esteban@cab.inta-csic.es}

\author[0000-0002-6985-9476]{F. M. Jim\'enez-Esteban}
\affiliation{Departmento de Astrof\'{\i}sica, Centro de Astrobiolog\'{\i}a (INTA-CSIC), ESAC Campus, Camino Bajo del Castillo s/n, E-28692 Villanueva de la Ca\~nada, Madrid, Spain}
\affiliation{Spanish Virtual Observatory, Madrid, Spain}
  
\author{E. Solano}
\affiliation{Departmento de Astrof\'{\i}sica, Centro de Astrobiolog\'{\i}a (INTA-CSIC), ESAC Campus, Camino Bajo del Castillo s/n, E-28692 Villanueva de la Ca\~nada, Madrid, Spain}
\affiliation{Spanish Virtual Observatory, Madrid, Spain}

\author{C. Rodrigo}
\affiliation{Departmento de Astrof\'{\i}sica, Centro de Astrobiolog\'{\i}a (INTA-CSIC), ESAC Campus, Camino Bajo del Castillo s/n, E-28692 Villanueva de la Ca\~nada, Madrid, Spain}
\affiliation{Spanish Virtual Observatory, Madrid, Spain}

\begin{abstract}

Binary and multiple stars have long provided an effective empirical method of testing stellar formation and evolution theories. In particular, the existence of wide binary systems (separations $>$\,20\,000 au) is particularly challenging to binary formation models as their physical separations are beyond the typical size of a collapsing cloud core ($\sim$\,5\,000\,--\,10\,000 au). We mined the recently published \GD\ catalog to identify bright comoving systems in the five-dimensional space (sky position, parallax, and proper motion). We identified 3\,741 comoving binary and multiple stellar candidate systems, out of which 575 have compatible RVs for all the members of the system. The candidate systems have separations between $\sim$\,400 and 500\,000 au. We used the analysis tools of the Virtual Observatory to characterize the comoving system members and to assess their reliability. The comparison with previous comoving systems catalogs obtained from TGAS showed that these catalogs contain a large number of false systems. In addition, we were not able to confirm the ultra-wide binary population presented in these catalogs. The robustness of our methodology is demonstrated by the identification of well known comoving star clusters and by the low contamination rate for comoving binary systems with projected physical separations $<$ 50\,000 au. These last constitute a reliable sample for further studies. The catalog is available online at the Spanish Virtual Observatory portal\footnote{http://svo2.cab.inta-csic.es/vocats/v2/comovingGaiaDR2/}.

\end{abstract}


\section{Introduction}

It is widely accepted that stars are born from the gravitational collapse of protostellar clouds. As it collapses, a molecular cloud breaks into smaller and smaller pieces in a hierarchical manner, until the fragments reach stellar mass. In this process, stars are not formed isolated but in embedded clusters, comprising up to a few thousand stars sharing roughly the same age and metallicity. \cite{Lada03} suggested that only a small fraction of the embedded clusters survive emergence from molecular clouds to become open clusters, associations of stars loosely bound by mutual gravitational attraction. Typically, after a few hundred million years, open clusters become disrupted by close encounters with other clusters and clouds of gas, as they orbit the Galactic center. As a remnant of this process, a significant fraction of main-sequence stars (the exact percentage depends on the spectral type) are in binary and multiple systems \citep[e.g.][]{Duquennoy91,Raghavan10}. 

Depending on authors, the definition of {\it wide binaries} is very ample, ranging from the minimum physical separation necessary to avoid Roche lobe filling and mass transfer after the main-sequence phase, to common proper motion pairs (pairs of stars traveling together through space with orbital periods ranging from a few thousand to millions of years and, thus, without any discernible relative orbital motion) with separations up to thousands of astronomical units \citep{Caballero09a}. Wide binaries are weakly bound and they can be disrupted by inhomogeneities in the Galactic potential due to stars or molecular clouds passing nearby. This makes these systems valuable indicators of the Galactic dynamical environment. Moreover, wide multiple systems can be regarded as mini star clusters with components born at the same time and with the same chemical composition, but evolving in an independent way due to the large physical separations among them. This makes them excellent candidates for testing stellar models and for age indicators.

There is not a consensus about the maximum physical separation at which a wide  pair  is  still  physically  bound  and  orbits  around  its  center  of  mass, and its numerical value has changed with time. During many years a cutoff at $\sim$\,20\,000 au  in projected physical separation of binaries was widely accepted  \citep[e.g.][]{Weinberg87,Close90}. However, \cite{Caballero09a} demonstrated the existence of multiple systems in the solar neighborhood with projected separations larger than this threshold. Such wide systems are more likely found in the Galactic halo, where the probability of encountering stars is minimal, and in young associations, where the systems have had less time for encounters. Recently, \cite{Oelkers17} claimed the discovery of the binary system with the largest separation, nearly 3.2 pc.

Over the past years, numerous wide binaries have been discovered (see, for instance, \citealt{Galvez-Ortiz17}). This type of object, with separations similar to or larger than the typical prestellar cores, poses a problem on classical models, in which the formation of binary systems is understood to occur during a coeval fragmentation of the giant molecular cloud. Various hypotheses have been proposed to overcome this matter. \cite{Reipurth12} suggest that triple systems are born very compact and can develop extreme hierarchical architectures on timescales of millions of years, as one component is dynamically scattered into a very distant orbit. The energy of ejection would come from shrinking the orbits of the other two stars. \cite{Kouwenhoven10}, on the other hand, suggested that these binary/multiple systems are not primordial, but they originate from different birth sites and become gravitationally bound during the dissolution phase of young star clusters. 

The first Gaia data release, {\it Gaia}-DR1 \citep{GaiaCollaboration16}, took place in September 2016 and included the astrometric dataset called TGAS (Tycho-Gaia Astrometric Solution), which contains positions, proper motions, and parallaxes for about 2 million of the brightest stars in the sky, which are in common with the {\it Hipparcos}  and Tycho-2 catalog. Three works were published with searches for wide binaries and moving groups in the TGAS catalog \citep{Andrews17,Oelkers17,Oh17}. The three of them utilized sophisticated Galactic models to identify comoving pairs using likelihood functions obtained from different statistical approaches. Each work identified several thousands of comoving candidate pairs with projected physical separations up to several parsecs. \cite{Oh17} and \cite{Oelkers17} found bimodal distributions for the projected separation and the binding energy, with a large number of pairs with separations $>$ 1 pc and energies below 10$^{-34}$ J. These authors explained the shape of the distributions in terms of two different binary populations: one of wide stable systems, corresponding to the higher energy and shorter separations peaks, which are expected to survive 10 Gyr or longer; and another one with ultra-wide young unstable systems, corresponding to the lower energy and larger separation peaks, which are not expected to last more than a few gigayears. The confirmation of a population of pairs with separations larger than 1 pc would provide further constraints in the Galactic gravitational potential. 

The second \G\ data release, \GD\ \citep{GaiaCollaboration18a}, took place in 2018 April. This release provides the astrometry solution for 1.3 billion sources, including those in TGAS. \GD\  uncertainties (typically $\sim$\,0.06 mas yr$^{-1}$ in proper motion and $\sim$\,0.04 mas in parallax) are better than the TGAS uncertainties by an order of magnitude for the parallax and two orders of magnitude for the proper motion. Furthermore, \GD\ includes new photometric filters ($G_{\rm BP}$ and $G_{\rm RP}$), and radial velocities (RVs) for several million stars with a mean $G$ magnitude between 4 and 13. All this makes \GD\ an ideal resource to search for comoving systems and to check the reliability of the results obtained with TGAS.

In this paper we describe the discovery and characterization of 3\,741  candidate pair and multiple systems found using the \GD\ catalog, and the comparison with the previous comoving systems from TGAS. The paper is organized in the following way: We describe the methodology used to search the comoving candidate systems in section \ref{method}; we present the catalog in section \ref{results}; we study the physical properties of the comoving candidate binary systems in section \ref{phys-prop}; we discuss our result in section \ref{discussion}; and, finally, we present our conclusions in section \ref{conclusion}.


\section{Search methodology}
\label{method}

\subsection{Search of comoving systems}

We used \GD\ data to search for comoving systems among the brightest stars in the sky, i.e. those included in the Tycho-2 catalog \citep{Hog00a}, using a simple approach based on the astrometric solution. Restricting our search to the $\sim$2.5 million Tycho-2 sources implies that comoving systems with a low-mass and faint companion will not be detected. Since the limiting magnitude of Tycho-2 was $\sim$11.5 mag in the visible, we estimate that Tycho-2 is completed up to $\sim$40 pc for M dwarfs, up to $\sim$200 pc for G dwarfs, and up to $\sim$400 pc for F dwarfs.

The search was carried out using a workflow consisting of the following steps:

\begin{itemize}

\item Our input file was the table {\it gaiadr2.tycho2\_best\_neighbour}, available in the Gaia archive at ESA\footnote{https://gea.esac.esa.int/archive/}. This table contains the list of \GD\ sources with Tycho-2 counterparts (hereafter the \GDT\ sample; \citealt{Marrese18}). Only sources with good astrometric solutions were considered. We followed the latest recommendations published by the \G\ ESA team in the {\it Known issues with the Gaia DR2 data} web page\footnote{https://www.cosmos.esa.int/web/gaia/dr2-known-issues}. According to them, we selected sources with the {\it re-normalised unit weight error} $RUWE<1.4$. (technical note GAIA-C3-TN-LU-LL-124-01\footnote{http://www.rssd.esa.int/doc\_fetch.php?id=3757412}). In addition, we restricted our sample to \GDT\ sources with positive parallaxes and relative errors $<$ 10\% in both proper motion components and parallax. The errors that we used were the external uncertainties calculated as indicated by \cite{Lindegren18b} at the IAU GA Division A meeting in Vienna on 2018 August 27\footnote{An extended version of this presentation can be found on the {\it Known issues with the Gaia DR2 data} web page.}. We used the correction proposed for \G\ objects fainter than 13 mag in $G$ band. Finally, given the Tycho-2 limiting magnitude, we removed any source with \G\ $G>13$.

The 10\% in the relative error criterion was conservative enough to keep pairs with small differences in the proper motion due to the orbital velocity and to avoid biases in the distance estimation due to large errors in parallax \citep[e.g.][]{Astraatmadja16}. This criterion, nevertheless, favored the selection of nearby objects with high proper motions, in the sense that it was more likely to find relative errors above 10\% in objects with low proper motion and/or parallax.  

This selection resulted in 1\,936\,422 sources (79\% of the total sources of {\it gaiadr2.tycho2\_best\_neighbour}). 

\item Systems with projected physical separations larger than 50\,000 au have been reported in the literature. \cite{Caballero10b} actually pushed this limit up to 1 pc for a multiple system in the Castor moving group. Other authors like \cite{Jiang10} supported the existence of wide binary stars with even larger separation from a theoretical point of view. Furthermore, the three TGAS catalogs of comoving systems aforementioned have a large number of pairs separated more than 1 pc. Thus, we decided to set a generous upper limit of 500\,000 au ($\sim$\,2.5 pc) assuming that pairs beyond this limit are most likely not physically bound. 

Therefore, for each of the 1\,936\,422 sources selected in the previous step, we looked for companions with the same parallax and proper motion in R.A. and decl. within 2.5$\sigma$ (typically 0.13 mas in parallax and 0.2 \masyr\ in proper motion), with $\sigma$ being the largest error of the two stars in the pair, and with a projected physical separation in the sky lower than 500\,000 au. Before, we applied the proper motion correction for bright stars due to inertial spin of the Gaia DR2 proper motion system \citep{Lindegren18b}. We used the classical expression to convert parallaxes to distances: $d$\,=\,1000/$\pi$, where $\pi$ is the parallax measured by Gaia in milliarcseconds and $d$ is the distance in parsecs. For each comoving pair candidates, the adopted parallax was the error weighted arithmetic mean. The distances derived from parallaxes and the small-angle approximation (tan$\rho$\,$\approx$\,$\rho$) were used to convert angular separations between components ($\rho$) into projected physical separation ($s$) using the formula $s$\,=\,$\rho$\,$\times$\,$d$.

\item In some cases, an individual source was assigned to more than one comoving companion, meaning that there were more than two sources sharing the same values (within the errors) of proper motion and parallax. These sources were considered to form multiple ($>$2 members) comoving systems. For multiple system {\it s} was calculated taken as reference the source closest to the center of the group, and using a weighted arithmetic mean parallax for all members of the system. This produced a number of cases where the projected physical separation between the central member and other members of the group was larger than the adopted threshold (500,000 au). However, for every source in the multiple system there was always another source separated less than the given threshold.

\end{itemize}

This workflow resulted in 11\,834 individual sources grouped into 3\,852 comoving binary and multiple candidate systems that fulfilled the previous conditions.

\subsection{Radial velocity rejection}
\label{RV}

If the members of our candidate systems are really physically bound they should present similar RVs within the errors. However, RVs of field stars have a typical dispersion of 20\,--\,40 km\,s$^{-1}$, depending on the direction of the sky. Thus, the probability of chance alignment within 2.5$\sigma$ (typically 2.3 \kms) is non-negligible. So, similar RVs cannot be used as a confirmation that two stars close in the sky form a physically bound system, but it is a good approach to find fake systems. We would like to stress that our candidate systems already share the same proper motion and distance. Therefore, coincident RVs is an additional indication that increases their chance of being real.

The \GD\ catalog provided RVs for 3\,318 of our sources. For the rest, we used TOPCAT\footnote{http://www.star.bris.ac.uk/$\sim$mbt/topcat/} to cross-match our candidate systems with the Radial Velocity Experiment (RAVE, DR5)\footnote{http://cdsarc.u-strasbg.fr/viz-bin/Cat?III/279/rave\_dr5} \citep{Kunder17} available at Vizier \citep{Ochsenbein00}. A 3\arcsec\ search radius was adopted after using the proper motion information to refer the position to the epoch J2000. We found 338 counterparts in the RAVE catalog. 

For our analysis, we were restrictive in the quality of the RV, using only values with errors lower than 5 km\,s$^{-1}$. This reduced the available data to 2\,877 from \GD\ and 282 from RAVE. Two hundred and eleven sources had RV data in both catalogs. In all but four cases, the RV values from \GD\ and RAVE were compatible within the errors. For these common sources we used \GD\ data because of their lower errors, typically better than 1 km\,s$^{-1}$. Thus, we collected RVs for 2\,948 individual sources of our candidate sample.

We compared the RVs of the members of each candidate system using the same criteria used for the proper motion and the parallax (values differing in less than 2.5$\sigma$). 678 candidate pairs had RV data available for both components. Of these, 108 pairs ($\sim$16\%) were found to have discrepant RVs. Moreover, 66 objects assigned to multiple system showed incompatible RVs with the rest of the members of the system, and they were consequently discarded. In total, we removed 282 sources on the basis of their RVs.

\cite{Makarov08} estimated a multiple hierarchical system rate larger than 25\% in a sample of comoving pairs within 25 pc from the Sun, while \cite{Tokovinin14} obtained a lower rate (13\%) in a sample of solar-like dwarfs within 67 pc from the Sun. Thus, part of the 16\% of candidate pairs with discrepant RV could be true wide binaries where one of the members has an unresolved companion, i.e., they could be multiple hierarchical high order systems.

\section{Results: comoving star catalog}
\label{results}

The final catalog of comoving candidate systems with the same (within the errors) parallaxes and proper motions in \GD\ is made up of 11\,550 sources grouped in 3\,741 systems out of which 575 have also the same RVs within the errors. 3\,055 systems are comoving pairs, but there are also 686 higher order multiple systems with up to 94 candidate members. Table\,\ref{tab:summary} summarizes the number of systems according to multiplicity.

All the relevant information of these systems can be gathered from {\em The SVO archive of \GD\ double and multiple bright stars} at the Spanish Virtual Observatory portal \footnote{http://svo2.cab.inta-csic.es/vocats/v2/comovingGaiaDR2/} (see the Appendix~\ref{Append}).

\begin{table}
  \caption{Number of systems according to multiplicity}
  \label{tab:summary}
  \begin{center}
  \begin{tabular}{rrr}
    \hline
    \hline
    \noalign{\smallskip}
Members & Systems & similar RV \\
    \noalign{\smallskip}    
\hline
    \noalign{\smallskip}                                           
2 & 3\,055 & 570 \\
3 &  288 &   4 \\
4 &  104 &     \\
5 &   63 &     \\
6 &   42 &   1 \\
7 &   34 &     \\
8 &   21 &     \\
9 &   16 &     \\
10 &  14 &     \\
$>$10$^{a}$ & 104 & \\
\noalign{\smallskip}                                            
    \hline                                                          
Total & 3\,741 & 575 \\
\noalign{\smallskip}                                            
    \hline                                                          
  \end{tabular}                                                     
  \begin{list}{}{}                                                  
\item[Notes:] 
$^{a}$ Candidate systems with more than 10 candidate members.
\end{list}
\end{center}                                                
\end{table}

\subsection{Spatial distribution of the comoving groups}

\begin{figure}
 \includegraphics[width=\textwidth]{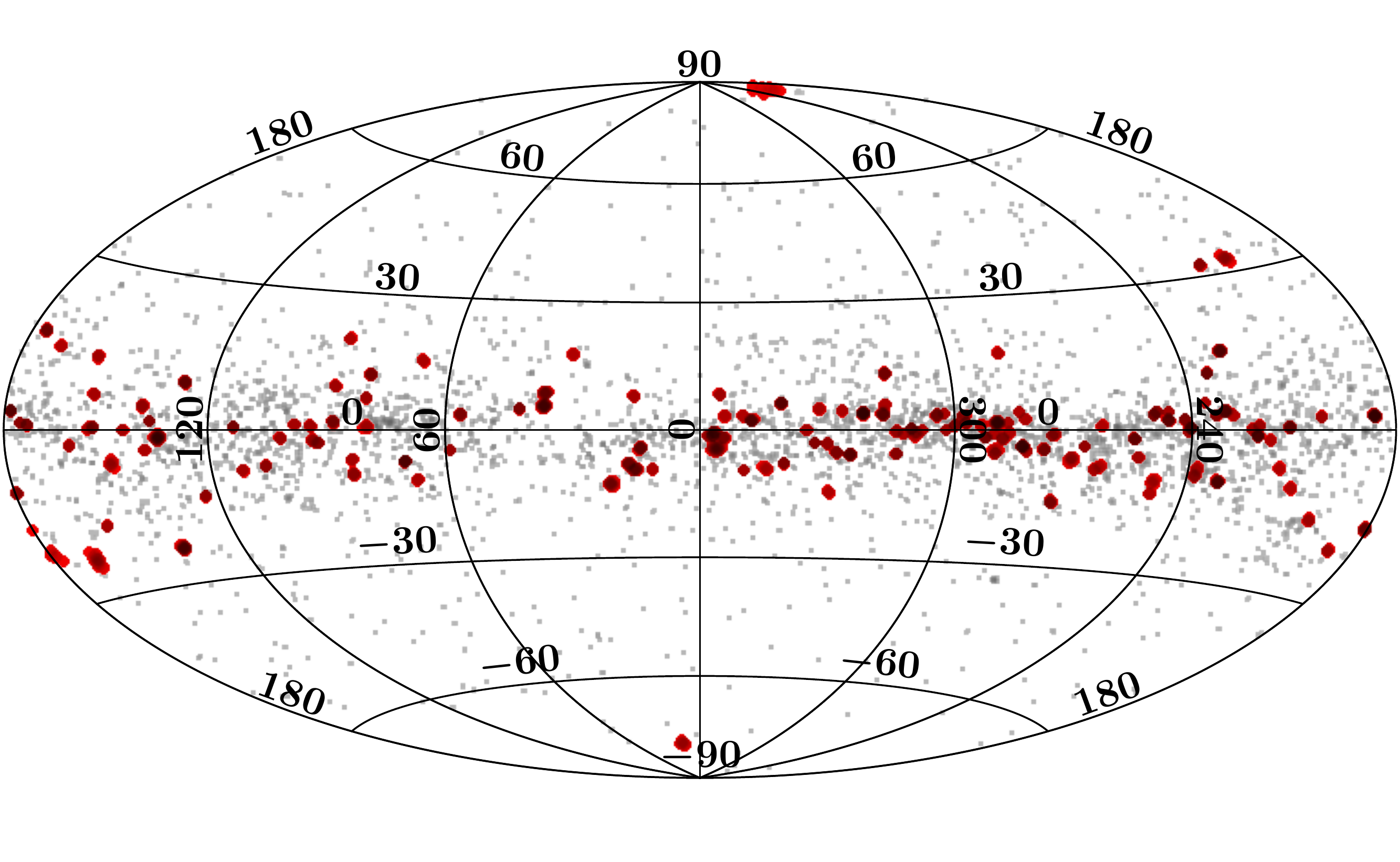}
  \caption{Sky distribution in Galactic coordinates of our comoving candidate sources. Filled red circles are the sources belonging to multiple systems with more than seven candidate members.}
  \label{fig:Skypos}
\end{figure}

Figure~\ref{fig:Skypos} shows the sky distribution of the sources of our comoving system catalog. Filled red circles represent the sources belonging to multiple systems with more than seven candidate members. There is a concentration in the direction of the Galactic plane, but avoiding the direction toward the Galactic bulge. This distribution is similar to the one presented in the Tycho double star catalog \citep[see Figure~2 in][]{Fabricius02}. The authors of this catalog explained it on the basis of the nature of the {\it Hipparcos}  scanning law. Since our search sample was constructed from Tycho-2, it is expected that our comoving systems present a similar distribution. 

As listed in Table\,\ref{tab:summary}, we found a large number of high order multiple systems, probably members of star clusters or moving groups, most of them concentrated in the Galactic plane. In fact, some of them belong to very well known star clusters (e.g. Melotte~22 or IC~2391), but we also found sources not reported in SIMBAD as cluster members.

A detailed analysis of the multiple systems is a demanding task that goes beyond the scope of this paper. Thus, in what follows, we will focus on the binary system candidates.

\section{Physical properties}
\label{phys-prop}

\subsection{Dwarf/Subgiant and giant separation}
\label{D/G}

\begin{figure*}
	\includegraphics[width=\textwidth]{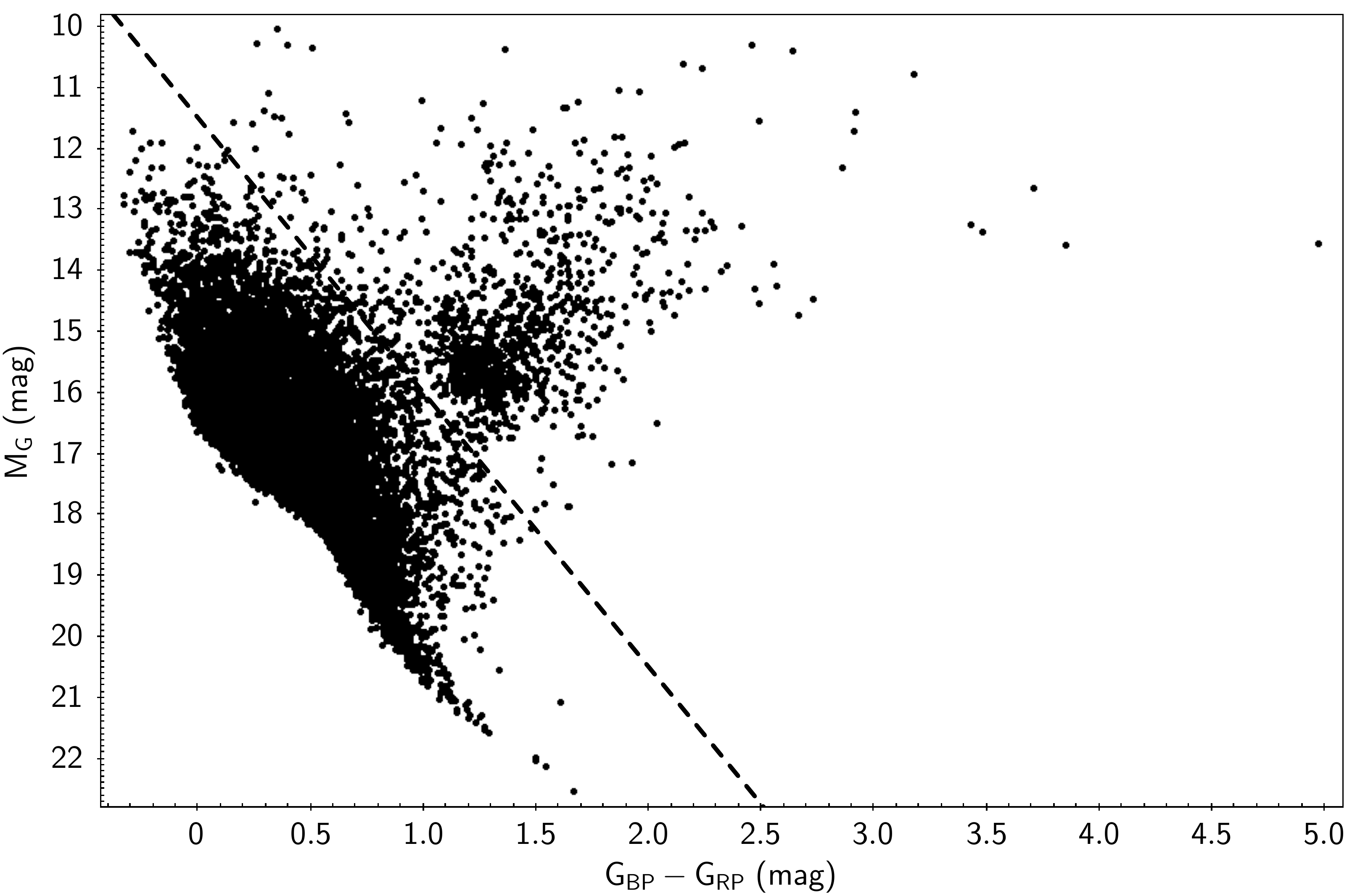}
    \caption{Hertzsprung--Russell diagram of our comoving candidate sample. The dashed line depicts the selected separation criterion between dwarfs/subgiants and giants (see the text).}
	\label{fig:HR}
\end{figure*}

We used the {\it Gaia} $M_{G}$ absolute magnitude ($M_{G}=G+5+5\log \pi$) versus {\it Gaia} $G_{\rm BP}-G_{\rm RP}$ color to build a Hertzsprung--Russell diagram (HRD). Figure\,\ref{fig:HR} shows the position in the HRD of our comoving candidate members.

We separated dwarfs/subgiants from giants by eye in the HRD considering dwarfs/subgiants all sources with $M_{G}$\,$\geq$\,4.5$\times (G_{\rm BP}-G_{\rm RP})$+11.5 (shown as a dashed line in Figure\,\ref{fig:HR}). According to their position in the HRD 1\,094 objects ($\sim$\,10\%) were classified as giants and 10\,458 as dwarfs/subgiants ($\sim$\,90\%).

\subsection{Effective temperatures}

The next step was the determination of effective temperatures ($T_{eff}$). Although \GD\ includes information on temperatures, these values, derived from the three broad photometric bands, may be affected by different issues \citep{Andrae18}. Furthermore, the estimation of $T_{eff}$ for \GD\ was restricted to the range from 3\,000 to 10\,000 K, which may not be the case for our sources. Thus, we decided to use VOSA\footnote{http://svo2.cab.inta-csic.es/theory/vosa/} \citep{Bayo08} to get our own estimation of the $T_{eff}$. VOSA is a Virtual Observatory tool designed to determine physical parameters from the comparison of observed photometry to different collections of theoretical models.

Using VOSA we queried the GALEX \citep{Bianchi00}, Tycho-2 \citep{Hog00a}, \GD\ \citep{GaiaCollaboration18a}, APASS (DR9; \citealt{Evans02}), 2MASS \citep{Skrutskie06} and {\it WISE} \citep{Wright10} photometric catalogs to build the Spectral Energy Distribution (SED) from the ultraviolet to the infrared. For those sources with no values of extinction found in VO archives, we used the ones provided by \GD. For the rest of the sources we assumed no extinction. Dereddened observational SEDs were then compared to the grid of BT-Settl model atmospheres \citep{Allard12}. We adopted  {\it log\,g}\,$\geq$\,4.0 and {\it log\,g}\,$<$\,4.0 for dwarfs/subgiants and giants (classified according to their position in Figure\,\ref{fig:HR}, see Sect.\,\ref{D/G}), respectively. We assumed solar metallicity for all of them. We used the VOSA internal goodness of fit criterion ($V_{gfb}$\footnote{Vgfb: Modified reduced $\chi^{2}$, calculated by forcing $\sigma(F_{obs})$ to be larger than $0.1\times F_{obs}$, where $\sigma(F_{obs})$ is the error in the observed flux ($F_{obs}$). This can be useful to avoid the risk of overweighting photometric points with underestimated photometric errors. $V_{gfb}$ smaller than 10--15 is often perceived as a good fit.} $<$15) to select the sources with good SED fitting (11,143 sources). 

\begin{figure}
	\includegraphics[width=\columnwidth]{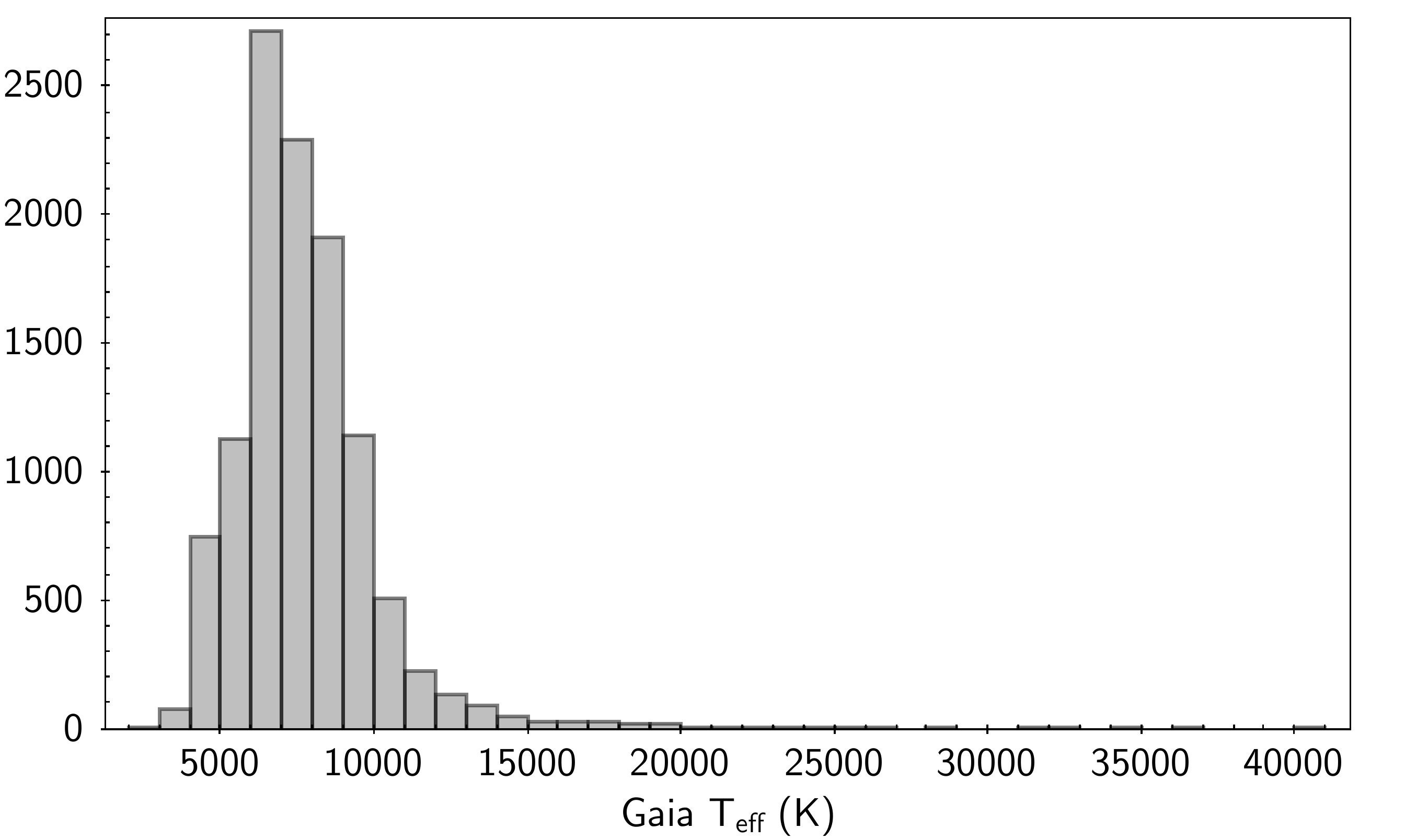}
    \caption{Effective temperature distribution of our sample.} 
	\label{fig:Hist_Teff}
\end{figure}

Figure\,\ref{fig:Hist_Teff} shows the distribution in effective temperature of our sample. Most of the sources have $T_{eff}$ between 3\,500 K (M5V) and 12\,500 K (B8V), with the maximum of the distribution at $\sim$6\,500\,K (F4V), as expected from the Tycho-2 limiting magnitude.

\begin{figure}
	\includegraphics[width=\columnwidth]{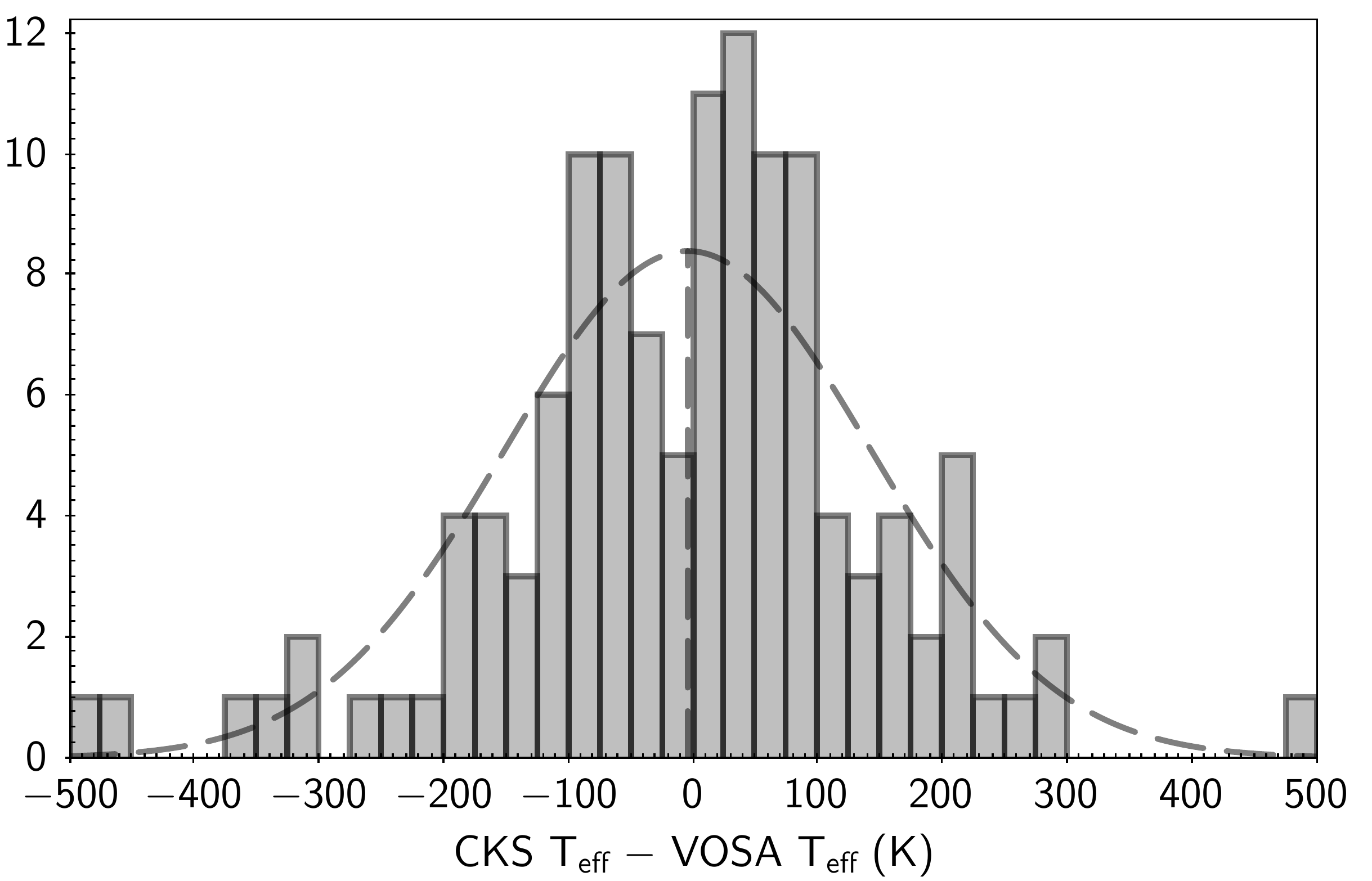}
    \caption{Comparison between the effective temperatures obtained with high-resolution spectroscopy by the CKS and with SED fitting by VOSA. The dashed line represents a Gaussian fit.}
	\label{fig:Diff_Teff}
\end{figure}

In order to evaluate the accuracy and reliability of the effective temperatures derived using the SED fitting method, we used a control sample consisting of 126 stars from the California-$Kepler$ Survey (CKS, \citealt{Petigura17}) included in the \GDT\ sample. \cite{Petigura17} used two different codes to obtain the physical parameters of these stars from high-resolution optical spectra, achieving an accuracy of 60 K in \Teff. We then used VOSA to construct the SEDs of these 126 stars and estimate the \Teff\ in an identical manner to the stars in our catalog. Figure\,\ref{fig:Diff_Teff} shows the comparison between the \Teff\, obtained from spectroscopy by the CKS and from the SEDs fitting by VOSA. The Gaussian fit gives a standard deviation of $\sim$147 K and shows that there is no systematic offset between both datasets. Assuming error propagation in quadrature and taking into account that the typical error in CKS is $\pm$60 K, we estimated the error in the effective temperatures calculated by VOSA in $\pm$135 K.

The existence of hierarchical tertiary unresolved companions that may have passed the RV filter could affect the estimation of the $T_{eff}$. However, for most of these cases, the combined SED will clearly deviate from the photospheric models and, therefore, the SED fitting will be poor. So it is expected that these systems did not fulfill our goodness of fit criterion ($V_{gfb}$ $<$ 15).

\subsection{Binding energies}

Stellar masses were derived from effective temperatures and the dwarf/subgiant and giant classification by interpolating in the Tables B1 and B2 of \cite{Gray08} for giants and dwarfs till M0. For cooler dwarfs we used the relationship given in \cite{Reid05}. Seventy-one giants had $T_{eff}$ beyond the range covered by \cite{Gray08} and their masses were not obtained. For each source with good SED fitting, we took 1\,$\sigma$ in $T_{eff}$ to obtain a minimum and a maximum $T_{eff}$, and used them to derive a minimum and a maximum masses. Finally, we assumed as the uncertainty in the mass estimation half of the difference between the minimum and maximum mass. From this exercise, we obtained a typical uncertainty in mass lower than 0.05\,\Msun. However, the assumption of solar metallicity in the SED fitting for all our stars introduced an additional uncertainty. With the aid of isochrones for stars of 5 Gyr age from the Dartmouth Stellar Evolution Database\footnote{http://stellar.dartmouth.edu/models/index.html} \citep{Dotter08}, we evaluated the effect of the metallicity in the mass estimation, obtaining an uncertainty of 0.1-0.2 \Msun\ depending on the \Teff\ of the star. Consequently, we assume that the typical mass uncertainty for our objects is typically lower than 0.2 \Msun.

Masses were then used to calculate the binding energy (U) of the comoving pairs, under the assumption that they are physical binaries. The binding energy of a pair is defined as the gravitational potential energy between the two objects {\it U\,=\,--GM$_{1}$M$_{2}$/a}, where {\it G} is the gravitational constant, {\it M$_{1}$} and {\it M$_{2}$} are the masses of each component and {\it a} represents the physical separation between the two stars. Since {\it a} is not a priori known, we use the logic of \cite{Fischer92} who assumed {\it a\,=\,1.26\,s}, where {\it s} is the projected physical separation in the plane of the sky. From the uncertainties in mass and parallax, we concluded that the typical uncertainty in the binding energies is $\sim$\,18\%.

\subsection{Main sequence and binary dissipation lifetimes}
\label{TdTms}

Even if a system possesses the necessary binding energy to remain stable, a large separation, coupled with the local Galactic environment, could cause the system to dissipate over time: encounters with other stars or molecular clouds or even subtle changes in the overall Galactic potential can contribute to disrupting the system's stability and cause it to break it apart \citep{Weinberg87}. Age is, thus, a fundamental parameter to accurately assess whether a binary system is really physically bound or not: the younger a pair is, the lower the probability of having had an encounter with stars or molecular clouds will be.

In this work we used the main-sequence lifetime ($t_{MS}$) as a proxy to estimate the ages of our objects. For the objects in the main sequence (the great majority of the members of our systems), $t_{MS}$ represents an upper limit in age, while for giants $t_{MS}$ is a lower limit. But even in this case, as the time spent by an object in the giant phase is relatively small compared to $t_{MS}$, this parameter can also be considered a reasonably good age estimator. Taking into account that the luminosity of a main-sequence star is typically assumed to be proportional to $M^{3.5}$, $t_{MS}$ can be estimated by using the following expression: 
\begin{equation}
t_{MS} \sim 10\,M/L = 10\,M^{-2.5}
\end{equation}
where $M$ and $L$ are the stellar mass and luminosity in solar units, and $t_{MS}$ is in gigayears.

Following \cite{Oelkers17} we define the dissipation lifetime ($t_{D}$) as:
\begin{equation}
t_{D} \sim 1.212\,M_{tot} / a
\end{equation}
where $a$ is the physical separation in parsecs, $M_{tot}$ is the total mass of the system in solar masses, and $t_{D}$ is in gigayears.

\begin{figure*}
  \centering
  \includegraphics[width=\textwidth]{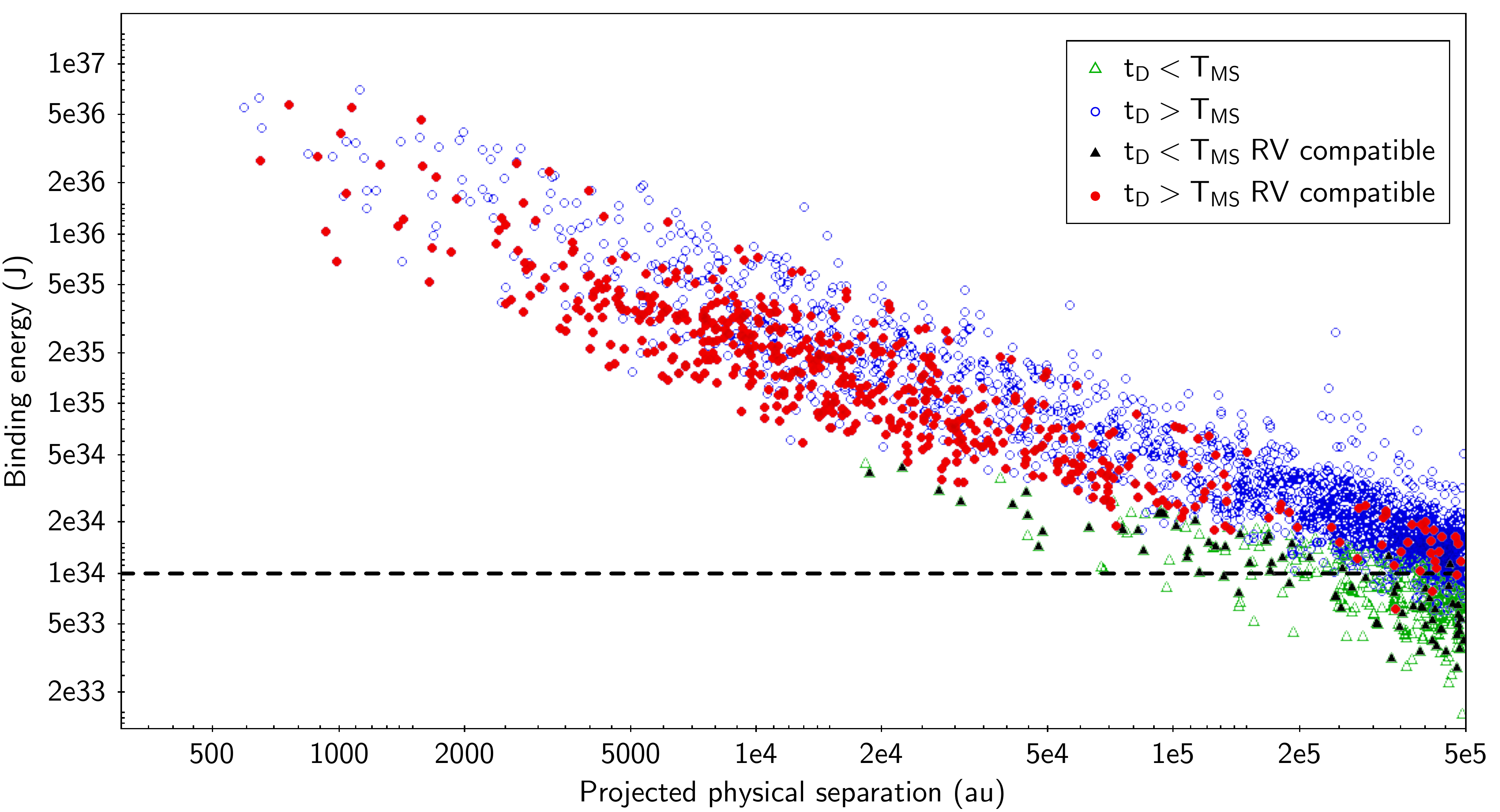}
  \caption{Binding energy as a function of the projected physical separation for the comoving candidate pairs of our catalog. Blue open circles represent systems with $t_{D}$\,$>$\,$t_{MS}$, and those with consistent RVs are plotted with red filled circles. Green open triangles show the systems with $t_{D}$\,$<$\,$t_{MS}$, and in black filled triangles  those with consistent RVs. The dashed horizontal line depicts the classical empirical limit for separation in wide binaries ($U_g$\,$\sim$\,10$^{34}$ J). Error bars are smaller than the symbol size.}
  \label{fig:s-U}
\end{figure*}

In Figure\,\ref{fig:s-U} we compare the dissipation and main-sequence lifetimes for the comoving candidate pairs of our catalog. The dashed horizontal line depicts the classical empirical limit of $U_g$\,=\,10$^{34}$ J for a system to be considered physically bound. The $t_{MS}$ of a pair was defined as that of the most massive member, i.e. the one with the shortest lifetime. We found 2\,359 candidate pairs with $t_{D}$\,$>$\,$t_{MS}$ (blue open circles), 558 of which had consistent RVs (red filled circles). Particularly interesting are the three systems (ID\,947, ID\,3145, and ID\,3331) with binding energies below 10$^{34}$ J and consistent RVs.

On the other hand, there are 429 candidate binary systems with $t_{D}$\,$<$\,$t_{MS}$ (green open triangles). Although these candidate systems have a high probability of being chance alignments, some of them, in particular, the youngest ones, can be real comoving systems. Among them, we can find some promising candidates based on the RV consistency (36 pairs with binding energies above 10$^{34}$ J and 41 below this limit, black filled triangles).

\section{Discussion}
\label{discussion}

\subsection{The contamination rate}
\label{cont-rate}

Our systems cover a large range in physical separations. We found comoving candidate pairs separated in the sky as close as only 1.8\arcsec\ up to as far as $\sim$1.4\arcdeg, and at distances from the Sun ranging from $\sim$40 to $\sim$2\,600 pc. This translates into a wide range of projected physical separations from $\sim$400 to 500\,000 au (the limit imposed in our search procedure).

Any search for comoving stars is affected by contamination by chance alignments (i.e., stars showing the same motion within the errors but that are actually not physically related). This contamination is expected to be stronger at large projected physical separations. The amount of contamination by chance alignments depends on the dataset used in the identification of the comoving systems, their nominal errors, and the selection criteria. Thus, in our case, it depends on the density of \GDT\ sources in a specific region of the sky, their astrometric solutions (parallaxes and proper motions, and their associated errors), and the adopted 2.5$\sigma$ criterion.

In order to assess the degree of contamination of our catalog, we followed two different approaches.

\subsubsection{Galaxy specular star test}

Following the exercise by \cite{Shaya11}, we took a star from the \GDT\ sample and we virtually moved it to the opposite side of the Galactic plane by multiplying by $-1$ its Galactic latitude. This way, we created a specular star in a region with similar stellar density and velocity field \citep{GaiaCollaboration18b}. Finally, we searched for comoving companions of the specular star using the criteria described in Sec\,\ref{method}. Any comoving systems found following this methodology were false positives due to change alignment.

\begin{figure*}
  \centering
  \includegraphics[width=\textwidth]{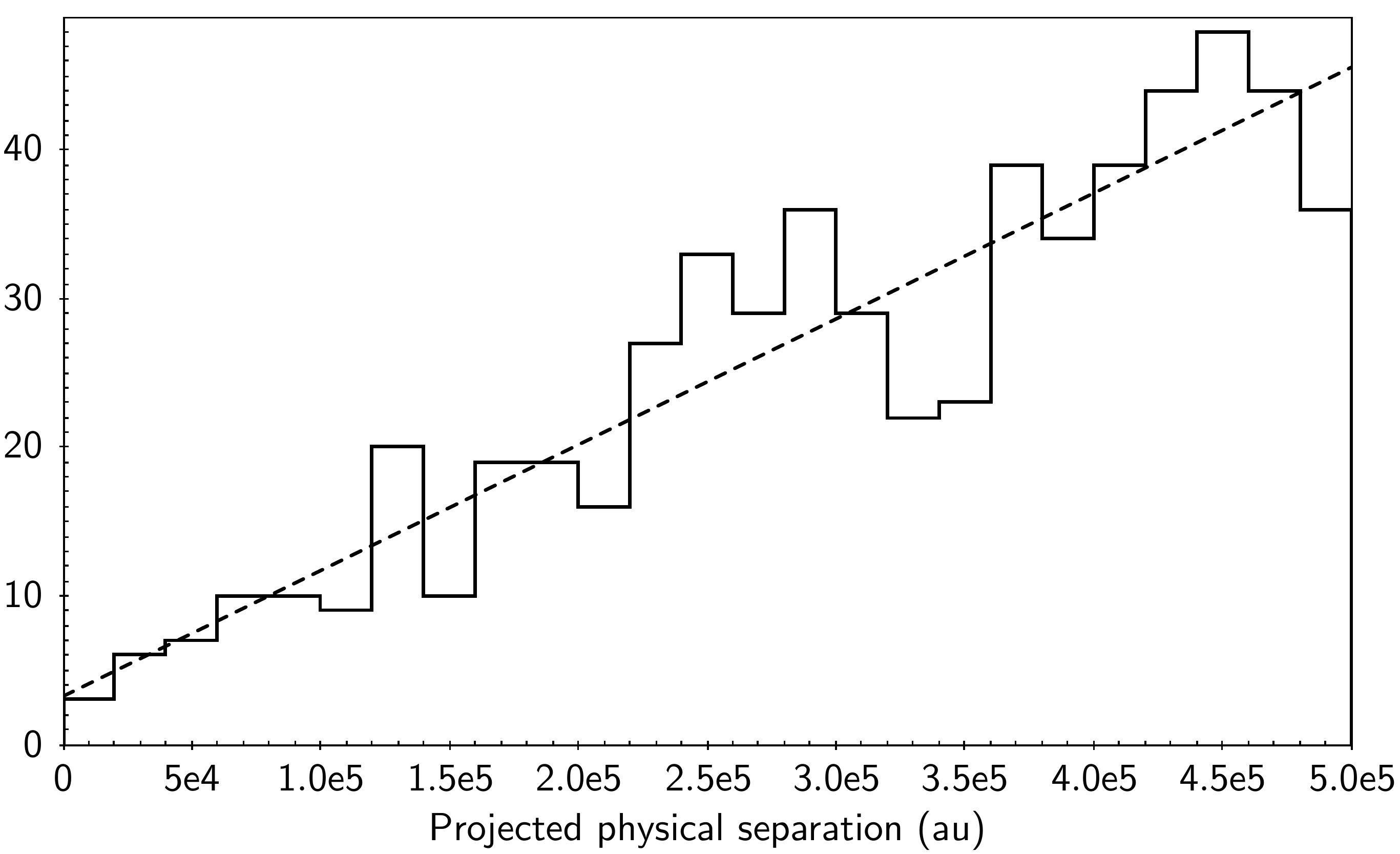}
  \caption{Distribution of the chance alignment false positives using the specular test as a function of the projected physical separation. The dashed line represents a linear fit.}
  \label{fig:FP-law}
\end{figure*}

After this exercise we obtained 714 false comoving systems: 617 false pairs and 97 high order false systems with up to 15 members. About 98\% of the false pairs had projected physical separation larger than 50\,000 au. Figure\,\ref{fig:FP-law} shows the distribution of the false comoving pairs in the function of the projected physical separation. It shows a clear correlation, increasing the number of false positives with the separation of the system.

Adopting this false positive occurrence as the expected in our catalog, it would imply that $\sim$20\% of our system would, indeed, be chance alignments. However, this ratio strongly depends on the projected physical separation being just $\sim$1\% of the candidate pairs with $s<$\,50\,000 au, $\sim$10\% for $s$ between 50\,000 and 100\,000 au, and increasing up to almost the 40\% for the largest separations. Thus, we should expect low contamination for small projected physical separations, especially below 50\,000 au.

\subsubsection{Expected chance alignment counterparts}

We also performed an exercise similar to that described in \cite{Smith18} to quantify the contamination rate associated with our comoving candidate binary systems. First we defined an inner circle with a radius corresponding to the projected physical separation of 500\,000 au at the pair's distance. This inner circle corresponds with the search area. Then, we defined an annulus constrained between the inner circle and an external radius five times larger than the inner circle radius, i.e. with an area 24 times larger than the searching area. We then searched for sources fulfilling our criteria in the annulus, and we assumed that all these matches were chance alignments. We calculated the number of expected false counterparts due to chance alignment within our search area multiplying the number of change alignment matches in the annulus ($M_{CA}$) by the ratio between the number of \GDT\ sources in the internal circle ($N_c$) and in the annulus ($N_a$). 

\begin{equation}
ECAC = M_{CA}\times N_c/N_a
\end{equation}

We refer to this number as the expected chance alignment counterparts (ECAC). It is important to note that ECAC is independent of the physical properties of the comoving candidate systems, like masses, separations, binding energies, as well as dissipation and main-sequence lifetimes.

For most of the comoving candidate pair members (4\,212 out of 6\,110), we did not find any matching source in the annulus, so ECAC was equal to zero. Sources with ECAC\,$>$\,0 ($\sim$ 31\% ) have a higher probability of being false positive due to chance alignments. Similarly to the previous exercise, they are evenly distributed with respect to projected physical separation, ranging from $\sim$5\% for $s<$\,50\,000 au, $\sim$18\% for $s$ between 50\,000 and 100\,000 au, and increasing up to to $\sim$51\% at largest separations. This reinforces our previous conclusion that most of the comoving candidate binaries with $s$\,$<$\,50\,000 au are real.

\subsection{Binding energy and projected physical separation distributions}
\label{distrib}

\begin{figure}
	\includegraphics[width=\columnwidth]{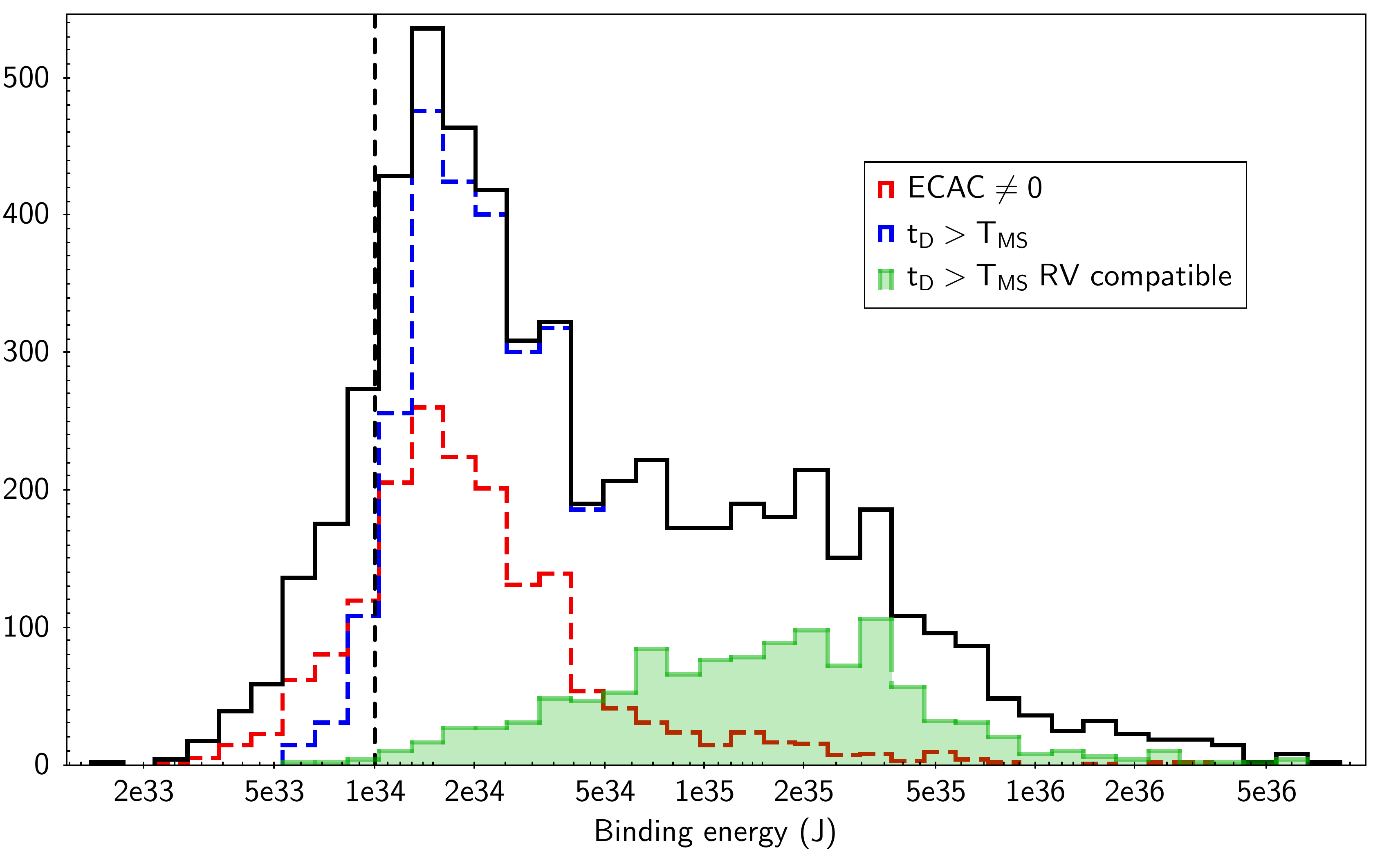}
    \includegraphics[width=\columnwidth]{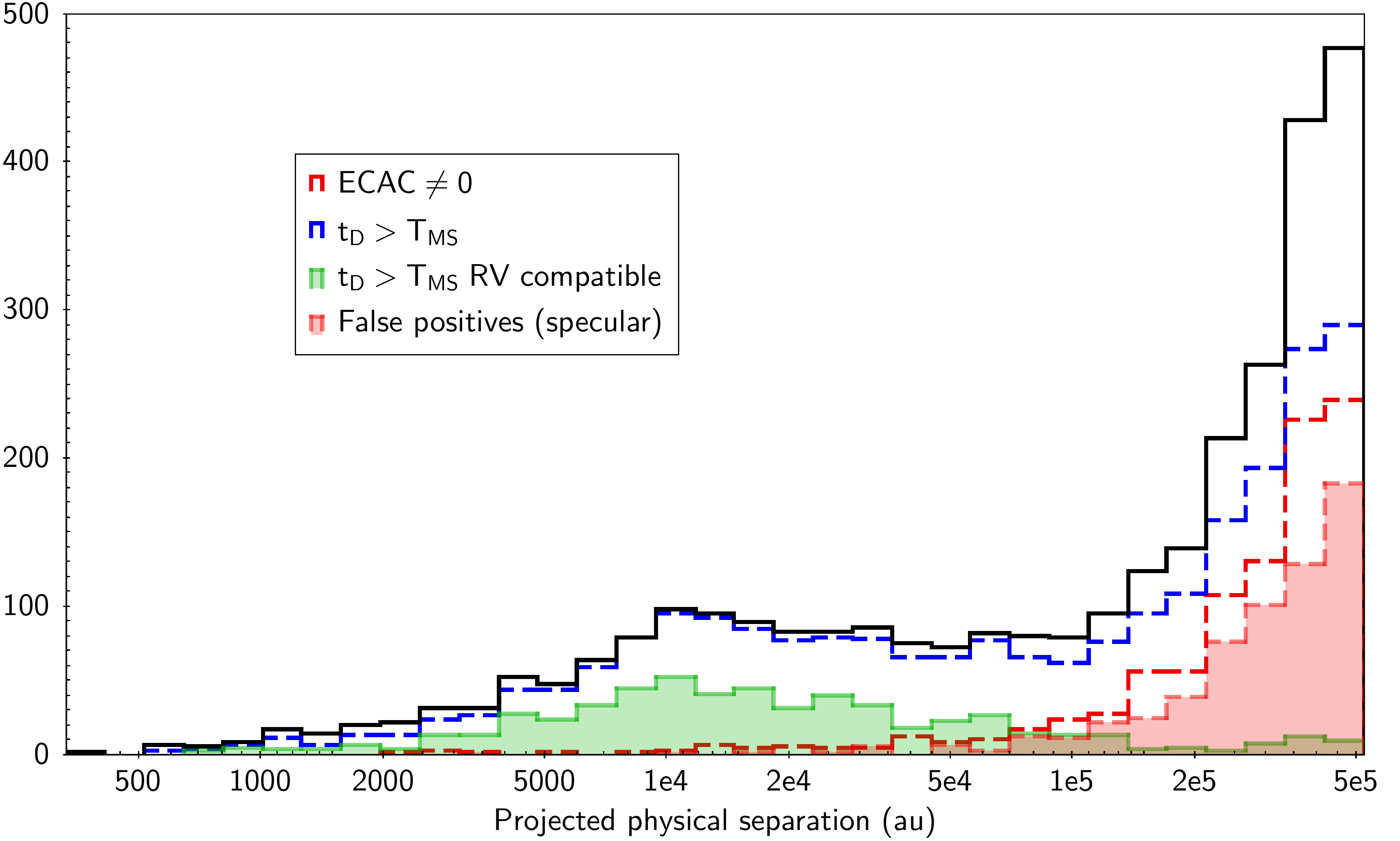}
    \caption{Binding energy (upper panel) and projected physical separation (lower panel) distributions of the pairs with calculated masses (black solid line). The blue dashed histogram represents the binary candidate systems with $T_D > T_{MS}$, while the filled green histogram shows the distribution of those fulfilling the previous condition and with consistent RVs. The red dashed line represents the binary candidate systems with $ECAC\neq$\,0 and the filled red histogram represents the chance alignment false positives (see Sect.\,\ref{cont-rate}). The vertical dashed line in the upper panel represents the classical empirical limit in binding energy.}
	\label{fig:U+S}
\end{figure}

Figure\,\ref{fig:U+S} shows the binding energy (upper panel) and projected physical separation (lower panel) distributions of the 2\,788 pairs with calculated masses (black solid line). In addition, we plotted the distribution of the subset of the systems with $t_{D}>t_{MS}$ (blue dashed line), and those last with compatible RV (filled green). Finally, we show the candidate systems with ECAC$\neq$0 (red dashed line), and the fake systems from the specular test (filled red at the bottom panel).

The binding energy distribution shows a clear bimodality, with all candidate pairs with $U$ above 10$^{33}$ J (10$^{40}$ erg), the lowest limit defined by \cite{Dhital10} and \cite{Caballero09a}, and most of them ($\sim$88\%) above the classical 10$^{34}$ J limit (vertical dashed line). As expected from the existent correlation between separation and binding energy (Figure\,\ref{fig:s-U}), the projected physical separation distribution also shows a bimodal shape, with a peak at $\sim$10\,000 au, then decreases up to $\sim$50\,000 au, and steeply increases up to 500\,000 au, the maximum separation imposed in this work. This bimodality is in agreement with the results found in SloWPoKes \citep{Dhital10,Dhital15} and, at least, two of the three catalogs based on TGAS data (see Sect.\,\ref{TGAS}). As already mentioned, these authors explained the shape of the distribution in terms of two different binary populations: one formed by wide stable systems (the higher energy and shorter separation peaks) and another with ultra-wide young unstable systems (the lower energy and longer separation peaks).

Nevertheless, as we report in Sec.\,\ref{cont-rate}, the number of false positives is expected to be high at large separations, and so, at low binding energies. This would imply that the peaks identified in these regions are mainly due to change alignment contamination rather than to real binary systems. Moreover, the bimodality disappears in both magnitudes if we restrict the analysis to sources with compatible RV. However, on the other hand, the expected number of false positives is not high enough to explain the whole distribution at large separations and low energies, implying that some of these pairs might be real binaries. Our conclusion is that although ultra-wide binaries may exist, they should be rare and they are probably binary systems that are dissipating. However, we have to wait for subsequent Gaia releases (Gaia-DR3 is foreseen for the first half of 2021) to confirm or reject this result on the basis of the RV information.

\subsection{Comparison with TGAS binary systems catalogs}
\label{TGAS}

\begin{figure}
  \centering
  \includegraphics[width=\columnwidth]{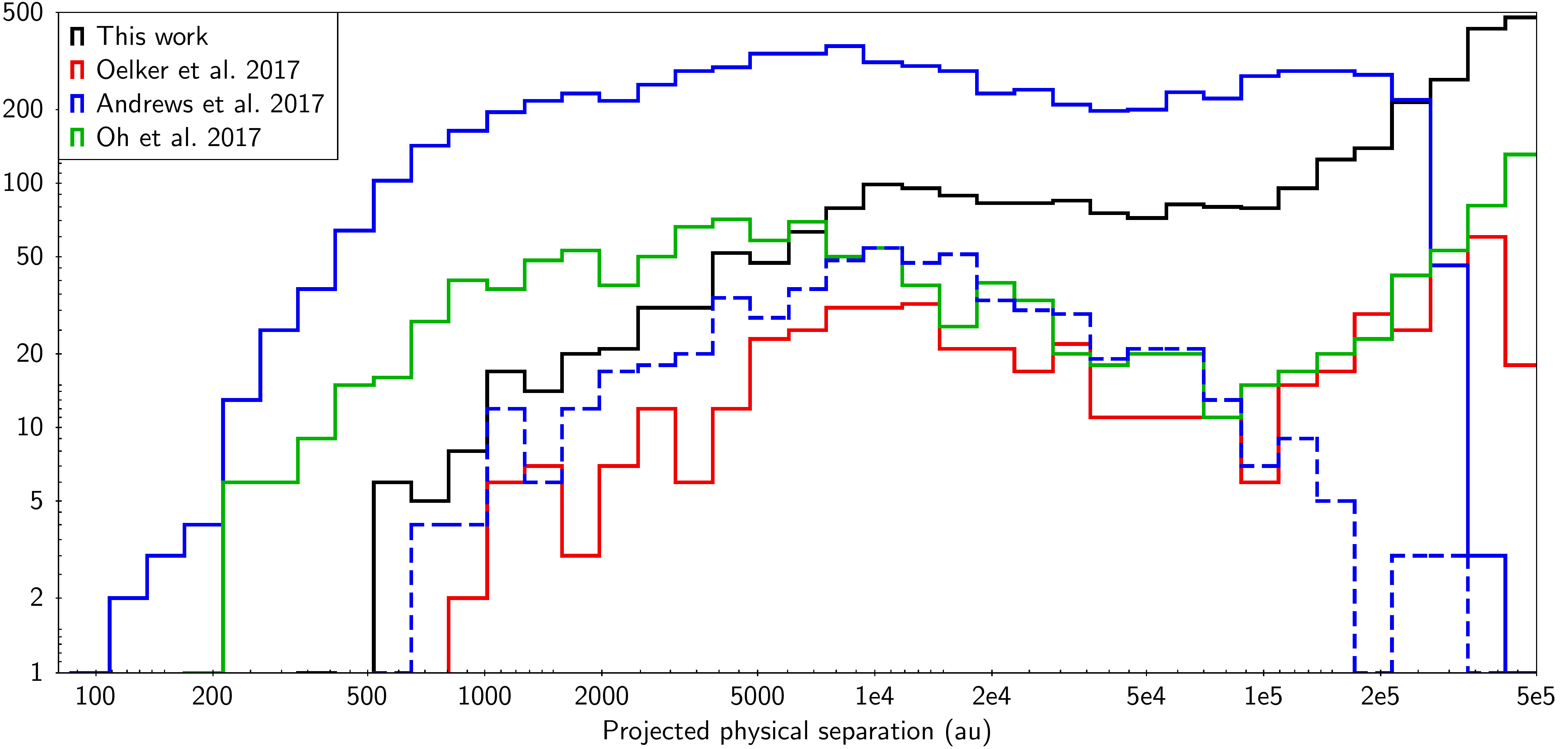}
  \caption{Comparison of the projected physical separation of our candidate binary systems with those obtained in the comoving TGAS catalogs. Solid lines represent the original distributions found, while the blue dashed line represents the candidate pairs in common among our sample and the sample by \cite{Andrews17}. We have truncated the distributions to $s$\,=\,500\,000 au, the limit imposed in this work. Colors are as labeled in the figure.}
  \label{fig:TGAS_comp}
\end{figure}

Our catalog of comoving systems was obtained from the subsample of \GD\ sources in common with the Tycho-2 catalog. Tycho-2 was the basis for the TGAS catalog. Thus, it is interesting to compare our results with those of the three comoving system catalogs obtained using TGAS data (the comoving TGAS catalogs, hereafter) and published in the last two years \citep{Andrews17,Oelkers17,Oh17}. Figure\,\ref{fig:TGAS_comp} shows the projected physical separation distributions of our candidate pairs and those of the comoving TGAS catalogs.

We then cross-matched our catalog with the three comoving TGAS catalogs. Only 34 out of 543, 82 out of 4\,236, and 601 out of 7\,108 of the systems reported in \cite{Oelkers17}, \cite{Oh17}, and \cite{Andrews17}, respectively, were in common with our catalog. The distribution of comoving candidate pairs in common with \cite{Andrews17} is shown in Figure\,\ref{fig:TGAS_comp} with a blue dashed line. This distribution does not show bimodality. This supports the previous conclusion that the peak at larger separations is probably due to contamination and not to a real population of ultra-wide binaries. 

The low percentage (2\,--\,8\%) of candidate pairs in common, in particular, with \cite{Oh17}, was completely unexpected. This is mainly because there is a high percentage of candidate binary systems based on TGAS data that do not pass the cuts in parallax and proper motion when the superior \GD\ astrometric solution is used. Thus, due to the lower accuracy of TGAS data, we would expect a significant number of nonphysical binary systems in the comoving TGAS catalogs. However, using \GD\ data, \cite{Andrews18} claimed that the level of contamination affecting \cite{Andrews17} was very low ($\sim$6\%), which seems not to support our findings.

\section{Conclusions}
\label{conclusion}
 
In this paper we searched for bright binary and multiple star systems using the Tycho-2 sources with information in the \GD\ catalog. We identified 11\,550 sources whose parallaxes, proper motions, and, whenever available, RVs are consistent with being physically bound systems. A total of 3\,471 comoving candidate systems were found, 3\,055 being binary systems. We also identified higher order comoving candidate systems, some of them already identified as members of Galactic open clusters or moving groups, but many others are reported for the first time in this work. 

The candidate system members are quite diverse in spectral type, ranging from O/B to M, although the majority of them have $T_{eff}$ corresponding to F and G spectral types as expected from the magnitude limited Tycho-2 catalog. The projected physical separation of the comoving candidate pairs ranges from $\sim$\,400 au to 500\,000 au, the maximum separation imposed in this work, with 1\,411 candidate binaries with projected physical separations larger than 1 pc.

The bimodal distribution previously reported in the literature is also seen in our analysis, both using projected physical separations and binding energies. We evaluated the contamination by chance alignment in our sample. The contamination rate was estimated to be very low for the systems with projected physical separation $s$\,$<$\,50\,000 au, which correspond to the higher binding energies. Thus, we are confident that these candidate systems constitute a reliable sample for further studies, like testing stellar evolution models or age calibration studies. However, the contamination by chance alignment increases at larger separations and lower energies. 

The comparison with other comoving system catalogs obtained using TGAS data offers two conclusions: i) TGAS-based catalogs are affected by a high degree of contamination of false candidate systems, most likely due to the lower accuracy of TGAS data; ii) the bimodality disappears, keeping only the peak at shorter physical separations and lower binding energies. Actually, the latter also happens if we restrict our analysis to the subsample of systems with RV information. Thus, we cannot confirm the ultra-wide binary population suggested in some of the comoving TGAS catalogs. Ultra-wide binary systems may exist, but they should be rare. However, it will be necessary to wait for the third \G\ data release, planned for the first half of 2021, to have a statistically significant number of systems with RV information to confirm or discard this conclusion.

\section*{Acknowledgements}
We thank to J.A. Caballero and M. Cort\'es-Contreras for their comments and useful discussions on an early draft of this paper. This work has made use of data from the European Space Agency (ESA) {\it Gaia} mission (\url{https://www.cosmos.esa.int/gaia}), processed by the {\it Gaia} Data Processing and Analysis Consortium (DPAC, \url{https://www.cosmos.esa.int/web/gaia/dpac/consortium}). Funding for the DPAC has been provided by national institutions, in particular the institutions participating in the {\it Gaia} Multilateral Agreement. This publication makes use of VOSA, developed under the Spanish Virtual Observatory project supported by the Spanish MINECO through grant AyA2017-84089. This research has made use of the VizieR catalog access tool, CDS, Strasbourg, France. We acknowledge use of the ADS bibliographic services. This research has made use of "Aladin sky atlas" developed at CDS, Strasbourg Observatory, France. This research has made use of Topcat \citep{Taylor05}. F.J.E. acknowledges financial support from the Spacetec-CM project (S2013/ICE-2822), and from ASTERICS project (ID:653477, H2020-EU.1.4.1.1. - Developing new world-class research infrastructures).

\software{Topcat \citep{Taylor05}, Aladin \citep{Bonnarel00}, VOSA \citep{Bayo08}}


\appendix
\section{Online catalog service}
\label{Append}

In order to help the astronomical community in the use of the catalog of comoving star systems, we have built an archive system that can be accessed from a web page\footnote{http://svo2.cab.inta-csic.es/vocats/v2/comovingGaiaDR2/} or through a Virtual Observatory ConeSearch\footnote{e.g. http://svo2.cab.inta-csic.es/vocats/v2/comovingGaiaDR2/cs.php?RA=301.708\&DEC=-67.482\&SR=0.1\&VERB=2}. Table\,\ref{tab:cat} lists the data accessible through the catalog.

\begin{table}
  \caption{Description of the Catalog}
  \label{tab:cat}
  \begin{center}
  \begin{tabular}{lll}
    \hline
    \hline
    \noalign{\smallskip}
Label & Unit & Description \\
    \noalign{\smallskip}    
\hline
    \noalign{\smallskip}                                           
Source\_GaiaDR2 & & {\it Gaia}-DR2 ID \\ 
Source\_Tycho2  & & Tycho-2 ID \\ 
GroupID & & ID for match group \\ 
GroupSize & & Number of rows in match group \\ 
Sep\_ang & arcsec & Angular separation in the plane of the sky \\ 
Sep\_sky & au & Projected physical separation in the plane of the sky \\ 
ra\_epoch2000 & deg & Right Ascension (J2000) \\ 
dec\_epoch2000 & deg & Declination (J2000) \\ 
ra & deg & {\it Gaia}-DR2 Right Ascension \\ 
dec & deg & {\it Gaia}-DR2 Declination \\ 
parallax & mas & {\it Gaia}-DR2 parallax \\    
parallax\_total\_error & mas &  Estimated {\it Gaia}-DR2 parallax total error (see the text)\\    
pmra\_corrected & mas\,yr$^{-1}$ & {\it Gaia}-DR2 proper motion in R.A. corrected for the inertial spin (see the text) \\    
pmra\_corrected\_error & mas\,yr$^{-1}$ & Error of the {\it Gaia}-DR2 proper motion in R.A. corrected for the inertial spin (see the text) \\    
pmdec\_corrected & mas\,yr$^{-1}$ & {\it Gaia}-DR2 proper motion in declination corrected for the inertial spin (see the text) \\    
pmdec\_corrected\_error & mas\,yr$^{-1}$ & Error of the {\it Gaia}-DR2 proper motion in declination corrected for the inertial spin (see the text) \\    
phot\_g\_mean\_mag & mag & {\it Gaia}-DR2 $G$ band magnitude \\    
phot\_bp\_mean\_mag & mag & {\it Gaia}-DR2 $G_{BP}$ band magnitude \\    
phot\_rp\_mean\_mag & mag & {\it Gaia}-DR2 $G_{RP}$ band magnitude \\    
RUWE &  & {\it Gaia}-DR2 re-normalized unit weight error \\    
radial\_velocity & km\,s$^{-1}$ & {\it Gaia}-DR2 radial velocity \\    
radial\_velocity\_error & km\,s$^{-1}$ & {\it Gaia}-DR2 radial velocity error \\    
HRV & km\,s$^{-1}$ & Heliocentric radial velocity (HRV) from RAVE \\    
HRV\_error & km\,s$^{-1}$ & Error on HRV (eHRV) from RAVE \\    
D/G &  & Dwarf/Giant classification (see the text) \\    
Teff & K & Effective temperature of the model in Kelvin from VOSA \\    
Teff\_error & K & Uncertainty in the effective temperature of the model in Kelvin from VOSA \\    
Av & mag & Av used in VOSA \\    
Mass & \Msun & Mass \\    
Mass\_error & \Msun & Mass error (see the text) \\    
U & J & Binding energy \\    
U\_error & J & Binding energy error (see the text) \\    
td & Gyr & Dissipation lifetime \\    
td\_error & Gyr & Dissipation lifetime error (see the text) \\    
tms & Gyr & Main-sequence lifetime \\    
tms\_error & Gyr & Main-sequence lifetime error (see the text) \\    
ECAC &  & Expected chance alignment counterparts \\    
 \noalign{\smallskip}                                            
    \hline                                                          
\noalign{\smallskip}                                            
    \hline                                                          
  \end{tabular}                                                     
\end{center}                                                
\end{table}

\subsection{Web access}

The archive system implements a very simple search interface that permits queries by coordinates and radius as well as by other parameters of interest. The user can also select the maximum number of sources to return (with values from 10 to unlimited).

The result of the query is an HTML table with all the sources found in the archive fulfilling the search criteria. The result can also be downloaded as a VOTable or a CSV file. Detailed information on the output fields can be obtained placing the mouse over the question mark (``?") located close to the name of the column. The archive also implements the SAMP\footnote{http://www.ivoa.net/documents/SAMP/} (Simple Application Messaging) Virtual Observatory protocol. SAMP allows Virtual Observatory applications to communicate with each other in a seamless and transparent manner for the user. This way, the results of a query can be easily transferred to other VO applications, such as, for instance, Topcat.
\bibliographystyle{aasjournal}
\bibliography{references} 

\begin{thebibliography}{}
\expandafter\ifx\csname natexlab\endcsname\relax\def\natexlab#1{#1}\fi

\bibitem[{{Allard} {et~al.}(2012){Allard}, {Homeier}, \& {Freytag}}]{Allard12}
{Allard}, F., {Homeier}, D., \& {Freytag}, B. 2012, Philosophical Transactions
  of the Royal Society of London Series A, 370, 2765

\bibitem[{{Andrae} {et~al.}(2018){Andrae}, {Fouesneau}, {Creevey}, {Ordenovic},
  {Mary}, {Burlacu}, {Chaoul}, {Jean-Antoine-Piccolo}, {Kordopatis}, {Korn},
  {Lebreton}, {Panem}, {Pichon}, {Th{\'e}venin}, {Walmsley}, \&
  {Bailer-Jones}}]{Andrae18}
{Andrae}, R., {Fouesneau}, M., {Creevey}, O., {et~al.} 2018, \aap, 616, A8

\bibitem[{{Andrews} {et~al.}(2017){Andrews}, {Chanam{\'e}}, \&
  {Ag{\"u}eros}}]{Andrews17}
{Andrews}, J.~J., {Chanam{\'e}}, J., \& {Ag{\"u}eros}, M.~A. 2017, \mnras, 472,
  675

\bibitem[{{Andrews} {et~al.}(2018){Andrews}, {Chanam{\'e}}, \&
  {Ag{\"u}eros}}]{Andrews18}
---. 2018, Research Notes of the American Astronomical Society, 2, 29

\bibitem[{{Astraatmadja} \& {Bailer-Jones}(2016)}]{Astraatmadja16}
{Astraatmadja}, T.~L., \& {Bailer-Jones}, C.~A.~L. 2016, \apj, 832, 137

\bibitem[{{Bayo} {et~al.}(2008){Bayo}, {Rodrigo}, {Barrado y Navascu{\'e}s},
  {Solano}, {Guti{\'e}rrez}, {Morales-Calder{\'o}n}, \& {Allard}}]{Bayo08}
{Bayo}, A., {Rodrigo}, C., {Barrado y Navascu{\'e}s}, D., {et~al.} 2008, \aap,
  492, 277

\bibitem[{{Bianchi} \& {GALEX Team}(2000)}]{Bianchi00}
{Bianchi}, L., \& {GALEX Team}. 2000, \memsai, 71, 1123

\bibitem[{{Bonnarel} {et~al.}(2000){Bonnarel}, {Fernique}, {Bienaym{\'e}},
  {Egret}, {Genova}, {Louys}, {Ochsenbein}, {Wenger}, \&
  {Bartlett}}]{Bonnarel00}
{Bonnarel}, F., {Fernique}, P., {Bienaym{\'e}}, O., {et~al.} 2000, \aaps, 143,
  33

\bibitem[{{Caballero}(2009)}]{Caballero09a}
{Caballero}, J.~A. 2009, \aap, 507, 251

\bibitem[{{Caballero}(2010)}]{Caballero10b}
---. 2010, \aap, 514, A98

\bibitem[{{Close} {et~al.}(1990){Close}, {Richer}, \& {Crabtree}}]{Close90}
{Close}, L.~M., {Richer}, H.~B., \& {Crabtree}, D.~R. 1990, \aj, 100, 1968

\bibitem[{{Dhital} {et~al.}(2010){Dhital}, {West}, {Stassun}, \&
  {Bochanski}}]{Dhital10}
{Dhital}, S., {West}, A.~A., {Stassun}, K.~G., \& {Bochanski}, J.~J. 2010, \aj,
  139, 2566

\bibitem[{{Dhital} {et~al.}(2015){Dhital}, {West}, {Stassun}, {Schluns}, \&
  {Massey}}]{Dhital15}
{Dhital}, S., {West}, A.~A., {Stassun}, K.~G., {Schluns}, K.~J., \& {Massey},
  A.~P. 2015, \aj, 150, 57

\bibitem[{{Dotter} {et~al.}(2008){Dotter}, {Chaboyer}, {Jevremovi{\'c}},
  {Kostov}, {Baron}, \& {Ferguson}}]{Dotter08}
{Dotter}, A., {Chaboyer}, B., {Jevremovi{\'c}}, D., {et~al.} 2008, \apjs, 178,
  89

\bibitem[{{Duquennoy} \& {Mayor}(1991)}]{Duquennoy91}
{Duquennoy}, A., \& {Mayor}, M. 1991, \aap, 248, 485

\bibitem[{{Evans} {et~al.}(2002){Evans}, {Irwin}, \& {Helmer}}]{Evans02}
{Evans}, D.~W., {Irwin}, M.~J., \& {Helmer}, L. 2002, \aap, 395, 347

\bibitem[{{Fabricius} {et~al.}(2002){Fabricius}, {H{\o}g}, {Makarov}, {Mason},
  {Wycoff}, \& {Urban}}]{Fabricius02}
{Fabricius}, C., {H{\o}g}, E., {Makarov}, V.~V., {et~al.} 2002, \aap, 384, 180

\bibitem[{{Fischer} \& {Marcy}(1992)}]{Fischer92}
{Fischer}, D.~A., \& {Marcy}, G.~W. 1992, \apj, 396, 178

\bibitem[{{Gaia Collaboration} {et~al.}(2016){Gaia Collaboration}, {Brown},
  {Vallenari}, {Prusti}, {de Bruijne}, {Mignard}, {Drimmel}, {Babusiaux},
  {Bailer-Jones}, {Bastian}, \& et~al.}]{GaiaCollaboration16}
{Gaia Collaboration}, {Brown}, A.~G.~A., {Vallenari}, A., {et~al.} 2016, \aap,
  595, A2

\bibitem[{{Gaia Collaboration} {et~al.}(2018{\natexlab{a}}){Gaia
  Collaboration}, {Katz}, {Antoja}, {Romero-G{\'o}mez}, {Drimmel}, {Reyl{\'e}},
  {Seabroke}, {Soubiran}, {Babusiaux}, {Di Matteo}, \&
  et~al.}]{GaiaCollaboration18b}
{Gaia Collaboration}, {Katz}, D., {Antoja}, T., {et~al.} 2018{\natexlab{a}},
  \aap, 616, A11

\bibitem[{{Gaia Collaboration} {et~al.}(2018{\natexlab{b}}){Gaia
  Collaboration}, {Brown}, {Vallenari}, {Prusti}, {de Bruijne}, {Babusiaux},
  {Bailer-Jones}, {Biermann}, {Evans}, {Eyer}, \&
  et~al.}]{GaiaCollaboration18a}
{Gaia Collaboration}, {Brown}, A.~G.~A., {Vallenari}, A., {et~al.}
  2018{\natexlab{b}}, \aap, 616, A1

\bibitem[{{G{\'a}lvez-Ortiz} {et~al.}(2017){G{\'a}lvez-Ortiz}, {Solano},
  {Lodieu}, \& {Aberasturi}}]{Galvez-Ortiz17}
{G{\'a}lvez-Ortiz}, M.~C., {Solano}, E., {Lodieu}, N., \& {Aberasturi}, M.
  2017, \mnras, 466, 2983

\bibitem[{{Gray}(2008)}]{Gray08}
{Gray}, D.~F. 2008, {The Observation and Analysis of Stellar Photospheres}

\bibitem[{{H{\o}g} {et~al.}(2000){H{\o}g}, {Fabricius}, {Makarov}, {Urban},
  {Corbin}, {Wycoff}, {Bastian}, {Schwekendiek}, \& {Wicenec}}]{Hog00a}
{H{\o}g}, E., {Fabricius}, C., {Makarov}, V.~V., {et~al.} 2000, \aap, 355, L27

\bibitem[{{Jiang} \& {Tremaine}(2010)}]{Jiang10}
{Jiang}, Y.-F., \& {Tremaine}, S. 2010, \mnras, 401, 977

\bibitem[{{Kouwenhoven} {et~al.}(2010){Kouwenhoven}, {Goodwin}, {Parker},
  {Davies}, {Malmberg}, \& {Kroupa}}]{Kouwenhoven10}
{Kouwenhoven}, M.~B.~N., {Goodwin}, S.~P., {Parker}, R.~J., {et~al.} 2010,
  \mnras, 404, 1835

\bibitem[{{Kunder} {et~al.}(2017){Kunder}, {Kordopatis}, {Steinmetz},
  {Zwitter}, {McMillan}, {Casagrande}, {Enke}, {Wojno}, {Valentini},
  {Chiappini}, {Matijevi{\v c}}, {Siviero}, {de Laverny}, {Recio-Blanco},
  {Bijaoui}, {Wyse}, {Binney}, {Grebel}, {Helmi}, {Jofre}, {Antoja}, {Gilmore},
  {Siebert}, {Famaey}, {Bienaym{\'e}}, {Gibson}, {Freeman}, {Navarro},
  {Munari}, {Seabroke}, {Anguiano}, {{\v Z}erjal}, {Minchev}, {Reid},
  {Bland-Hawthorn}, {Kos}, {Sharma}, {Watson}, {Parker}, {Scholz}, {Burton},
  {Cass}, {Hartley}, {Fiegert}, {Stupar}, {Ritter}, {Hawkins}, {Gerhard},
  {Chaplin}, {Davies}, {Elsworth}, {Lund}, {Miglio}, \& {Mosser}}]{Kunder17}
{Kunder}, A., {Kordopatis}, G., {Steinmetz}, M., {et~al.} 2017, \aj, 153, 75

\bibitem[{{Lada} \& {Lada}(2003)}]{Lada03}
{Lada}, C.~J., \& {Lada}, E.~A. 2003, \araa, 41, 57

\bibitem[{{Lindegren} {et~al.}(2018){Lindegren}, {Hernandez}, {Bombrun},
  {Klioner}, {Bastian}, {Ramos-Lerate}, {de Torres}, {Steidelmuller},
  {Stephenson}, {Hobbs}, {Lammers}, \& {Biermann}}]{Lindegren18b}
{Lindegren}, L., {Hernandez}, J., {Bombrun}, A., {et~al.} 2018, in , XXXth
  General Assembly of the IAU Symposium

\bibitem[{{Makarov} {et~al.}(2008){Makarov}, {Zacharias}, \&
  {Hennessy}}]{Makarov08}
{Makarov}, V.~V., {Zacharias}, N., \& {Hennessy}, G.~S. 2008, \apj, 687, 566

\bibitem[{{Marrese} {et~al.}(2018){Marrese}, {Marinoni}, {Fabrizio}, \&
  {Altavilla}}]{Marrese18}
{Marrese}, P.~M., {Marinoni}, S., {Fabrizio}, M., \& {Altavilla}, G. 2018,
  ArXiv e-prints, arXiv:1808.09151

\bibitem[{{Ochsenbein} {et~al.}(2000){Ochsenbein}, {Bauer}, \&
  {Marcout}}]{Ochsenbein00}
{Ochsenbein}, F., {Bauer}, P., \& {Marcout}, J. 2000, \aaps, 143, 23

\bibitem[{{Oelkers} {et~al.}(2017){Oelkers}, {Stassun}, \&
  {Dhital}}]{Oelkers17}
{Oelkers}, R.~J., {Stassun}, K.~G., \& {Dhital}, S. 2017, \aj, 153, 259

\bibitem[{{Oh} {et~al.}(2017){Oh}, {Price-Whelan}, {Hogg}, {Morton}, \&
  {Spergel}}]{Oh17}
{Oh}, S., {Price-Whelan}, A.~M., {Hogg}, D.~W., {Morton}, T.~D., \& {Spergel},
  D.~N. 2017, \aj, 153, 257

\bibitem[{{Petigura} {et~al.}(2017){Petigura}, {Howard}, {Marcy}, {Johnson},
  {Isaacson}, {Cargile}, {Hebb}, {Fulton}, {Weiss}, {Morton}, {Winn}, {Rogers},
  {Sinukoff}, {Hirsch}, \& {Crossfield}}]{Petigura17}
{Petigura}, E.~A., {Howard}, A.~W., {Marcy}, G.~W., {et~al.} 2017, \aj, 154,
  107

\bibitem[{{Raghavan} {et~al.}(2010){Raghavan}, {McAlister}, {Henry}, {Latham},
  {Marcy}, {Mason}, {Gies}, {White}, \& {ten Brummelaar}}]{Raghavan10}
{Raghavan}, D., {McAlister}, H.~A., {Henry}, T.~J., {et~al.} 2010, \apjs, 190,
  1

\bibitem[{{Reid} \& {Hawley}(2005)}]{Reid05}
{Reid}, I.~N., \& {Hawley}, S.~L. 2005, {New light on dark stars : red dwarfs,
  low-mass stars, brown dwarfs}, doi:10.1007/3-540-27610-6

\bibitem[{{Reipurth} \& {Mikkola}(2012)}]{Reipurth12}
{Reipurth}, B., \& {Mikkola}, S. 2012, \nat, 492, 221

\bibitem[{{Shaya} \& {Olling}(2011)}]{Shaya11}
{Shaya}, E.~J., \& {Olling}, R.~P. 2011, \apjs, 192, 2

\bibitem[{{Skrutskie} {et~al.}(2006){Skrutskie}, {Cutri}, {Stiening},
  {Weinberg}, {Schneider}, {Carpenter}, {Beichman}, {Capps}, {Chester},
  {Elias}, {Huchra}, {Liebert}, {Lonsdale}, {Monet}, {Price}, {Seitzer},
  {Jarrett}, {Kirkpatrick}, {Gizis}, {Howard}, {Evans}, {Fowler}, {Fullmer},
  {Hurt}, {Light}, {Kopan}, {Marsh}, {McCallon}, {Tam}, {Van Dyk}, \&
  {Wheelock}}]{Skrutskie06}
{Skrutskie}, M.~F., {Cutri}, R.~M., {Stiening}, R., {et~al.} 2006, \aj, 131,
  1163

\bibitem[{{Smith} {et~al.}(2018){Smith}, {Lucas}, {Kurtev}, {Smart}, {Minniti},
  {Borissova}, {Jones}, {Zhang}, {Marocco}, {Contreras Pe{\~n}a}, {Gromadzki},
  {Kuhn}, {Drew}, {Pinfield}, \& {Bedin}}]{Smith18}
{Smith}, L.~C., {Lucas}, P.~W., {Kurtev}, R., {et~al.} 2018, \mnras, 474, 1826

\bibitem[{{Taylor}(2005)}]{Taylor05}
{Taylor}, M.~B. 2005, in Astronomical Society of the Pacific Conference Series,
  Vol. 347, Astronomical Data Analysis Software and Systems XIV, ed.
  P.~{Shopbell}, M.~{Britton}, \& R.~{Ebert}, 29

\bibitem[{{Tokovinin}(2014)}]{Tokovinin14}
{Tokovinin}, A. 2014, \aj, 147, 87

\bibitem[{{Weinberg} {et~al.}(1987){Weinberg}, {Shapiro}, \&
  {Wasserman}}]{Weinberg87}
{Weinberg}, M.~D., {Shapiro}, S.~L., \& {Wasserman}, I. 1987, \apj, 312, 367

\bibitem[{{Wright} {et~al.}(2010){Wright}, {Eisenhardt}, {Mainzer}, {Ressler},
  {Cutri}, {Jarrett}, {Kirkpatrick}, {Padgett}, {McMillan}, {Skrutskie},
  {Stanford}, {Cohen}, {Walker}, {Mather}, {Leisawitz}, {Gautier}, {McLean},
  {Benford}, {Lonsdale}, {Blain}, {Mendez}, {Irace}, {Duval}, {Liu}, {Royer},
  {Heinrichsen}, {Howard}, {Shannon}, {Kendall}, {Walsh}, {Larsen}, {Cardon},
  {Schick}, {Schwalm}, {Abid}, {Fabinsky}, {Naes}, \& {Tsai}}]{Wright10}
{Wright}, E.~L., {Eisenhardt}, P.~R.~M., {Mainzer}, A.~K., {et~al.} 2010, \aj,
  140, 1868

\end{thebibliography}

\end{document}